\documentclass[useAMS,usenatbib]{mn2e}
\usepackage{graphicx}
\usepackage{epsfig}
\usepackage{color}
\usepackage{tabularx}
\usepackage{multirow}
\usepackage[para,online,flushleft]{threeparttable}

\title[{\it Suzaku} observation of Kes 69]{A {\it Suzaku} X-ray study of the mixed-morphology supernova remnant Kes 69 and searching for its gamma-ray counterpart}
\author[Sezer et al.]{A.~Sezer,$^{1}$\thanks{E-mail: aytap.sezer@avrasya.edu.tr (AS)}
T.~Ergin$^{2}$\thanks{ergin.tulun@gmail.com (TE)}, R.~Yamazaki$^{3}$, Y.~Ohira$^{4}$ and N.~Cesur$^{5}$\\
$^{1}$Department of Electrical-Electronics Engineering, Avrasya University, 61250, Trabzon, Turkey\\
$^{2}$TUBITAK Space Technologies Research Institute, ODTU Campus, 06800, Ankara, Turkey\\
$^{3}$Department of Physics and Mathematics, Aoyama Gakuin University, 5-10-1 Fuchinobe, Sagamihara 252-5258, Japan\\
$^{4}$Department of Earth and Planetary Science, The University of Tokyo, 7-3-1 Hongo, Bunkyo-ku, Tokyo 113-0033, Japan \\
$^{5}$Department of Physics, Y{\i}ld{\i}z Technical University, 34220, Istanbul, Turkey
}
\begin{document}
\date{}
\pagerange{\pageref{firstpage}--\pageref{lastpage}} \pubyear{2018}
\maketitle
\label{firstpage}

\begin{abstract}

Kes 69 is a mixed-morphology (MM) supernova remnant (SNR) that is known to be interacting with molecular clouds based on 1720 MHz hydroxyl (OH) maser emission observations in the northeastern and southeastern regions. We present an investigation of Kes 69 using $\sim$67 ks {\it Suzaku} observation. The X-ray spectrum of the whole SNR is well fitted by a non-equilibrium ionization model with an electron temperature of $kT_{\rm e}$ $\sim$ 2.5 keV, ionization time-scale of $\tau$ $\sim$ 4.1$\times10^{10}$ cm$^{-3}$ s and absorbing column density of $N_{\rm H}$ $\sim$ 3.1$\times10^{22}$ cm$^{-2}$. We clearly detected the Fe-K$\alpha$ line at $\sim$6.5 keV in the spectra. The plasma shows slightly enhanced abundances of Mg, Si, S and Fe indicating that the plasma is likely to be of ejecta origin. We find no significant feature of a recombining plasma in this SNR. In order to characterize radial variations in the X-ray spectral parameters, we also analyze annular regions in the remnant. We investigate the explosive origin of Kes 69 and favor the core-collapse origin. Additionally, we report a lack of significant gamma-ray emission from Kes 69, after analyzing the GeV gamma-ray data taken for about 9 years by the Large Area Telescope on board {\it Fermi}. Finally, we discuss the properties of Kes 69 in the context of other interacting MM SNRs.

\end{abstract}


\section{Introduction}
Galactic supernova remnant (SNR) Kesteven 69 (also known as G21.8$-$0.6, hereafter Kes 69) has an extended incomplete radio shell with an angular diameter of $\sim$20 arcmin (e.g. \citealt{Sh70}). {\it ROSAT} and {\it Einstein} observations showed that this SNR has an irregular X-ray morphology, which shows correlation with an incomplete radio shell \citep{Se90, Yu03}. Based on the bright X-ray emission from the interior of the radio shell and a possible thermal X-ray spectrum, this remnant was classified as a mixed-morphology (MM) SNR \citep{Yu03}. \citet{Gr97} detected a compact OH maser at velocity of $\sim$69.3 km s$^{-1}$ in the northeastern part of the remnant using the Very Large Array (VLA) and Australia Telescope Compact Array (ATCA) observations. At $\sim$85 km~s$^{-1}$ velocity, \citet{He08} detected extended OH maser emission toward the southern bright radio shell in the Green Bank Telescope OH maser survey. Millimeter band observations of CO and HCO$^{+}$ lines toward Kes 69 provided strong evidence of an association between SNR and the $\sim$$+$85 km s$^{-1}$ component of molecular gas \citep{Zh09}. {\it Spitzer} observations of the remnant revealed the bright molecular emission lines of OH, CO and H$_{2}$, excited by a shock \citep{Re06, He09a}.

X-ray observations of Kes 69 were carried out by {\it Einstein}, {\it ROSAT}, {\it ASCA} and {\it XMM-Newton} \citep{Se90, Yu03, Bo12, Se13}. \citet{Yu03} used {\it ROSAT} Position Sensitive Proportional Counter data and found that the thermal X-ray spectrum is well fitted by VMEKAL model with an absorbing column density of $N_{\rm H}$ $\sim$ 2.4 $\times10^{22}$ cm$^{-2}$ and an electron temperature of $kT_{\rm e}$ $\sim$ 1.6 keV. \citet{Bo12} searched for compact hard X-ray sources in the field of Kes 69 using {\it XMM-Newton} data and reported on the detection of 18 hard X-ray sources in the 3.0$-$10.0 keV. \citet{Se13} analyzed {\it XMM-Newton} data and showed that the plasma of Kes 69 has a collisional ionization equilibrium (CIE) condition and yielded a plasma temperature of $kT_{\rm e}$ $\sim$ 0.62 keV and a column absorption of $N_{\rm H}$ $\sim$ 2.85 $\times10^{22}$ cm$^{-2}$. 

The kinematic distance to Kes 69 is estimated as $\sim$11.2 kpc \citep{Gr97} from the OH maser velocity, $\sim$5.5 kpc from H\,{\sc i} and $^{13}$CO spectra \citep{Ti08}, and $\sim$5.2 kpc from the SNR/molecular cloud (MC) association \citep{Zh09}. Assuming the Sedov stage solution, \citet{Bo12} estimated an age of the SNR about 10, 000 yr from {\it XMM-Newton} data.  \citet{Se13} derived the age of the remnant to be $\sim$$2.5\times10^{4}$ yr assuming a distance of 5.5 kpc.

Kes 69 was searched in TeV and GeV gamma-ray bands by H.E.S.S. and the Large Area Telescope detector on board {\it Fermi} Gamma-Ray Space Telescope ({\it Fermi}-LAT), respectively. In the 1st {\it Fermi}-LAT Supernova Remnant Catalog \citep{Ac16}, the upper limits on the flux of this SNR were reported on Table 3, i.e. among the not-detected SNRs, and H.E.S.S. missed Kes 69 in TeV gamma rays \citep{Bo11}. Recently, although a preliminary analysis of {\it Fermi}-LAT data showed the detection of GeV gamma rays from Kes 69, \citet{Er16} cautioned that there are many bright gamma-ray sources in the close neighborhood of Kes 69 that could be contributing to this result. 

Several SNRs are known as MM SNRs, which are identified by shell emission in the radio and centrally brightened thermal emission in the X-ray band with little or no shell brightening \citep{Rh98, La06, Vi12}. However, the nature of the centrally brightened X-ray emission is poorly understood. The overionized/recombining plasma (RP) was observed in many MM SNRs (e.g. \citealt{Ya09, Oz09, Er17, Su18}). In these SNRs, the ionization temperature ($kT_{\rm z}$) has been found to be significantly higher than the electron temperature ($kT_{\rm e}$). Most MM SNRs exhibit the GeV/TeV gamma-ray emission (e.g., \citealt{Ab09, Ab10a, Ca10}). They are associated with MCs as indicated by CO line emission and/or OH (1720 MHz) masers. There is a strong association between maser-emitting (ME) and MM SNRs, which implies that MM SNRs require dense molecular gas in their environment to show the unusual centrally filled X-ray morphology (e.g., \citealt{Sl17, Yu03, Wh91}). Kes 69 has a bright X-ray emission from the interior of the radio shell and interacting with MC based on 1720 MHz OH maser emission observations; thus, Kes 69 is considered to be a member of ME-MM SNRs \citep{Yu03}. 

In this paper, we study the high spectral resolution {\it Suzaku} observation of Kes 69 to investigate the nature of the X-ray emission and address the observed spectral characteristics. Additionally, we search for the gamma-ray counterpart of Kes 69 using archival {\it Fermi}-LAT data. The rest of the paper is organized as follows: We first summarize observations and data reduction in Section 2. In Section 3, we describe our analysis. Then, in Section 4, we discuss the nature of the thermal X-ray emission with a comparison to previous X-ray studies (Section 4.1), gamma-ray results (Section 4.2), its properties in the context of other interacting MM SNRs (Section 4.3) and explosion origin of Kes 69 (Section 4.4). The conclusions are summarized in Section 5.

\section{Observations and Data Reduction}
\subsection{X-ray Data}
Kes 69 was observed with the {\it Suzaku} X-ray Imaging Spectrometer (XIS: \citealt{Ko07a}) on 2014 September 27-29 (ObsID: 509037010). The net exposure of the cleaned event data was $\sim$67.2 ks. The XIS consists of four charge-coupled device (CCD) cameras placed at the focus of the X-Ray Telescope  (XRT: \citealt{Se07}), covering the 0.2$-$12 keV energy range. Each XIS sensor has 1024$\times$1024 pixels and covers a 17.8$\times$17.8 arcmin$^2$ field of view (FoV). The XIS 0, 2, and 3 are front side illuminated CCDs, whereas XIS1 is a back-side illuminated CCD. Since 2006 November 9, the XIS2 has not been available for observations. We retrieved {\it Suzaku} archival data of Kes 69 from the Data Archives and Transmission System (DARTS)\footnote{http://www.darts.isas.jaxa.jp/astro/suzaku/}.
Data reduction and analysis were made with {\sc Headas} software version 6.20\footnote{https://heasarc.nasa.gov/lheasoft/} and {\sc xspec} version 12.9.1 \citep{Ar96} with AtomDB v3.0.9\footnote{http://www.atomdb.org} \citep{Sm01,Fo12}.  We used a cleaned event file created by the {\it Suzaku} team, combined the 5$\times$5 and 3$\times$3 editing mode event files using {\sc xselect} v2.4d and generated the Redistribution Matrix Files and Ancillary Response Files using {\sc xisrmfgen} and {\sc xissimarfgen} tools \citep{Is07}. All spectra are binned to a minimum of 25 counts per bin to allow use of the $\chi^{2}$ statistic using the ftool {\sc grppha}.

\subsection{Gamma-ray Data}
{\it Fermi}-LAT data obtained between 2008-08-04 and 2017-03-07 were used in this analysis. We selected the {\it Fermi}-LAT Pass 8 `Source' class and front$+$back type events coming from zenith angles smaller than 90$^{\circ}$ and from a circular region of interest (ROI) with a radius of 30$^{\circ}$ centred at the SNR's radio position using \texttt{gtselect} of Fermi Science Tools (FST). 

The exposure map produced by the likelihood calculations gives the total exposure (in terms of effective area multiplied by time) for a given position on the sky producing counts inside the ROI. The time that {\it Fermi}-LAT observed at a given position on the sky at a given off-axis angle is called the livetime and it is used to calculate the exposure map. The FST \texttt{gtltcube} tool produces an array of livetimes at all points on the sky and saves it into a livetime cube. Then the \texttt{gtexpcube2} tool of FST can be used to apply the livetime cube to the ROI to generate a binned exposure map. In our analysis, the livetime calculated over the whole sky ranged between 26 mins and 388 hours and 14 exposure maps were produced for the analysis region of 15$^{\circ}$ $\times$  15$^{\circ}$, where each of the maps corresponds to a different logarithmic energy bin assigned between 200 MeV and 300 GeV.

\section{Analysis}
\subsection{X-ray Spectral Analysis}
In Figure 1, we present the XIS1 image of Kes 69 in the 0.3$-$10.0 keV energy band. The radio data from National Radio Astronomy Observations (NRAO) Very Large Array (VLA) Sky Survey (NVSS: \citealt{Co98}) are overlaid for comparison. The cross shows the position of maser emission at velocity $\sim$69.3 km s$^{-1}$ \citep{Gr97}. The spectral extraction regions are shown with the circles, while the eliminated hard point-like sources detected by {\it XMM-Newton} \citep{Bo12} are shown with magenta circles (radius 0.33 arcmin). The calibration sources at the corner of the CCD chips are excluded from the image.

\subsubsection{Background Estimation}
For the background analysis of {\it XMM-Newton} data, \citet{Bo12} used an empty area in the northeastern part of the FoV as a background region.  We extracted the background spectra from the FoV of the XIS, shown by the dashed area in Figure 1, excluding the calibration sources at the corners of the FoV, the point-like sources, and the emission from the remnant.  

Recent studies have demonstrated that an accurate estimation of the X-ray background is particularly important to generate the X-ray spectrum of extended sources such as SNRs (e.g., G346.6$-$0.2: \citealt{Ya13}; G32.8$-$0.1: \citealt{Ba16a}). For example, the flux of the Galactic ridge X-ray emission (GRXE) strongly depends on the location of the SNRs (e.g., \citealt{Uc13, YaS16}). We therefore consider the following background components: the non-X-ray background (NXB) as an instrumental background component and the GRXE and cosmic X-ray background (CXB) components as the X-ray background. We generated the NXB using {\sc xisnxbgen} \citep{Ta08} and subtracted it from the extracted spectrum. Then, we fitted the NXB-subtracted background spectra with the following model:

\begin{eqnarray}
{Abs_{\rm CXB} \times (power-law)_{\rm CXB} + Abs_{\rm GRXE} \times (apec + apec)}
\label{eqn:pi0}
\end{eqnarray}
where the apec is a CIE plasma model in the {\sc xspec}. The second term is the GRXE component. We assumed the CXB spectrum as a power-law of photon index 1.4 and CXB having a surface brightness of 5.4$\times$10$^{-15}$ erg s$^{-1}$ cm$^{-2}$ arcmin$^{-2}$ in the 2$-$10 keV energy band \citep{Ku02}. The X-ray background spectrum was simulated by using the {\sc fakeit} command in {\sc xspec} and then it was subtracted from the observed spectrum.

\subsubsection{Spectral Fits}
\begin{figure*}
\centering \vspace*{1pt}
\includegraphics[width=0.53\textwidth]{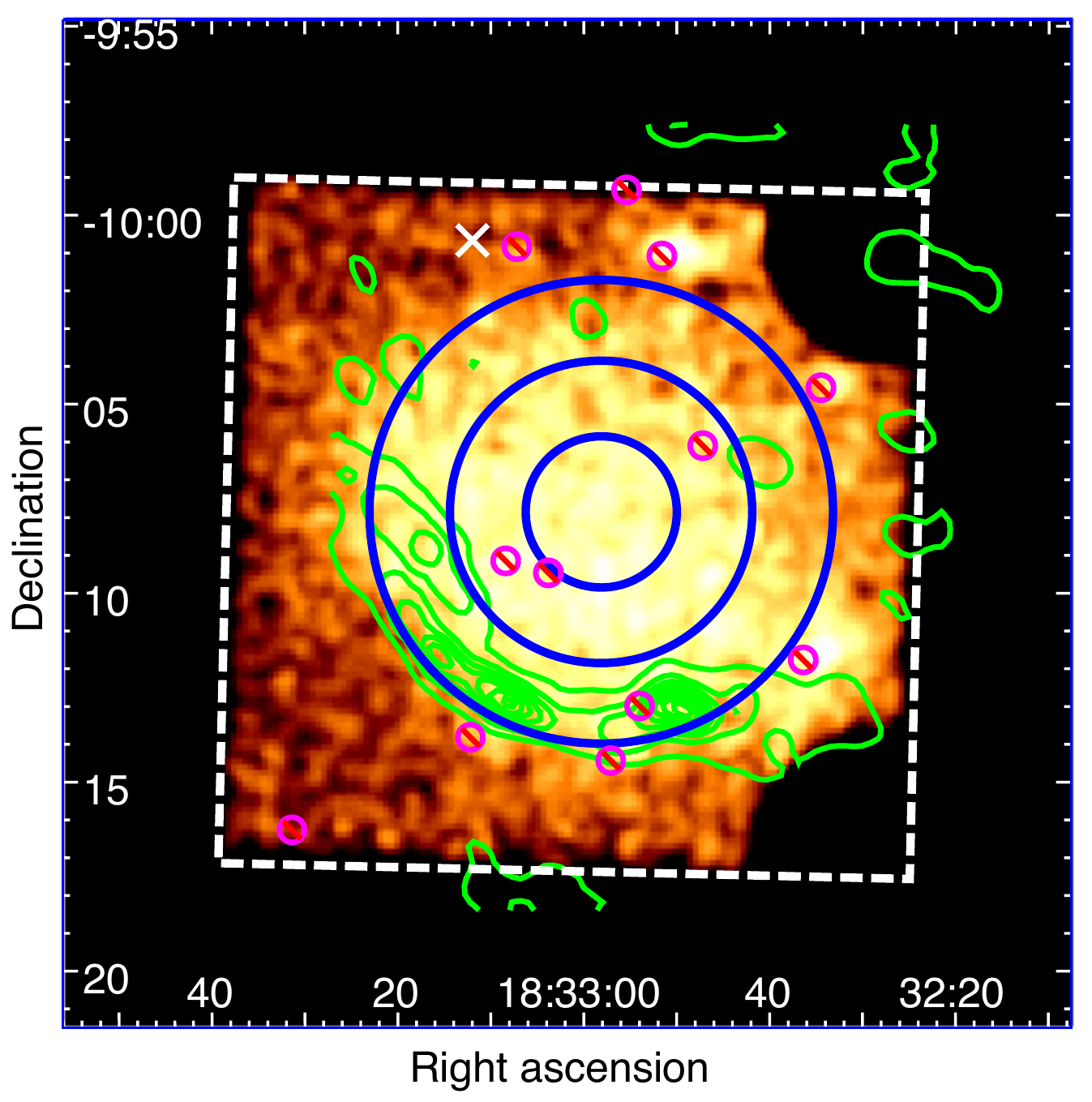}
\caption{{\it Suzaku} XIS1 image of Kes 69 in the 0.3$-$10.0 keV energy band, overlaid with the NRAO VLA radio contours \citep{Co98}. The radio contour levels are 0.005, 0.02, 0.08, 0.12 and 0.16 mJy beam$^{-1}$. The source regions are shown by the solid circles. Small magenta circles show the excluded point-like sources detected by {\it XMM-Newton} \citep{Bo12}. The XIS FoV is shown by the dashed area. We extracted background region from the entire XIS FoV, excluding the calibration regions and the source regions. The cross indicates the direction of the compact OH (1720 MHz) maser at the velocity of $\sim$69.3 km s$^{-1}$ associated with Kes 69 \citep{Gr97}. In the image, north is up and east is to the left.}
\label{figure_1}
\end{figure*}

In order to characterize the X-ray emitting plasma of the whole SNR, we first extracted the XIS spectra from the circular region with a radius of $\sim$6.1 arcmin, which is shown in Figure 1. To search for spectral variations in the temperature, ionization state and elemental abundances of Kes 69, we also extracted spectra from three annular regions of 0$-$2, 2$-$4 and 4$-$6 arcmin centred around $\rmn{RA}(J2000)=18^{\rmn{h}} 32^{\rmn{m}} 58^{\rmn{s}}$, $\rmn{Decl.}~(J2000)=-10\degr 07\arcmin 51\arcsec$, which are regions 1, 2 and 3, respectively, as shown in Figure 1.

The previous observations suggested that the X-ray spectrum of Kes 69 is characterized by a thermal plasma in CIE \citep{Yu03, Bo12, Se13}. Therefore, we first applied an absorbed (TBABS: \citealt{Wi00}) single-component variable-abundance CIE model (VMEKAL model in {\sc xspec}). The absorbing column density $N_{\rm H}$, electron temperature $kT_{\rm e}$, abundances of Mg, Si, S and Fe, and normalization were allowed to vary freely during the fitting. The other metal abundances were fixed to the solar values \citep{Wi00}. In this case, the electron temperature $kT_{\rm e}$ was obtained to be $\sim$1.2 keV, but this model was statistically unacceptable (with a reduced-$\chi^{2}$ $>$ 1.4).

As a next step, we employed an absorbed non-equilibrium ionization (NEI) plasma model (VNEI model in {\sc xspec}), where the free parameters were the interstellar absorption, $N_{\rm H}$, electron temperature, $kT_{\rm e}$, ionization time-scale, $n_{\rm e}t$, where $n_{\rm e}$ and $t$ represent the electron density and the time after the shock heating, respectively. The abundances of Mg, Si, S and Fe were also free parameters, while the other abundances were fixed to the solar values. We obtained a statistically acceptable fit (with a reduced-$\chi^{2}$ value of 1.19) for the whole region. To confirm the spectral parameters, we fitted the spectra with a single-component plane-parallel shock model with variable abundances ({\sc xspec} model VPSHOCK; \citealt{Bo01}) modified by interstellar absorption. We obtained an equally good fit with similar parameters to VNEI.

In order to check whether a cool component is needed across the remnant, we applied the two-component thermal plasma model (VNEI+VMEKAL) to the spectrum. For VNEI component, abundances of Mg, Si, S and Fe were free parameters. The abundances of all elements were assumed to the solar values for VMEKAL component. This model slightly improved the fit ($\chi^{2}_{\nu}$/dof=1.13/1539), but it did not give well-constrained spectral parameters. We therefore employed one-temperature NEI plasma model to fit the spectra of our spectral regions and concluded that the VNEI model represents the spectra. The background-subtracted 0.9$-$8.0 keV spectra are presented in Figure 2. For the annulus regions, we fixed the absorption $N_{\rm H}$ to that obtained for the whole SNR and then fitted each spectrum to a VNEI model. We obtain statistically acceptable fits (with reduced-$\chi^{2}$ values of 1.04$-$1.13). In Table 1, we present the best-fitting parameters with 90 per cent confidence ranges for all regions. 

We also tried an absorbed RP model (VRNEI\footnote{https://heasarc.gsfc.nasa.gov/xanadu/xspec/manual/node210.html} in {\sc xspec}) to search for the RP. This model is characterized by final electron temperature ($kT_{\rm e}$) and initial electron temperature ($kT_{\rm init}$), elemental abundances and a single ionization time-scale ($\tau$). We find that an initial plasma temperature ($\sim$0.57 keV) much smaller than the current plasma temperature ($\sim$2.59 keV) indicating that the plasma of this SNR is under-ionized. We also obtain $kT_{\rm init}$ $<$ $kT_{\rm e}$ for our annular regions indicating that the plasma is still ionizing.

\begin{table*}
 \begin{minipage}{170mm}
  \caption{Best-fitting spectral parameters with 90 per cent confidence ranges.}
 \begin{threeparttable}
\renewcommand{\arraystretch}{1.5}
  \begin{tabular}{@{}p{1.7cm}p{4cm}p{1.7cm}p{2.1cm}p{2.1cm}p{2.1cm}@{}}
  \hline\hline
     Component& Parameters & Whole& Region 1 &Region 2 &Region 3\\
\hline
TBABS & $N_{\rm H}$ ($10^{22}$ cm$^{-2})$ & $3.1_{-0.2}^{+0.2}$          & 3.1 (fixed)              & 3.1 (fixed)              & 3.1 (fixed) \\

VNEI & $kT_{\rm e}$ (keV)&  $2.5_{-0.2}^{+0.4}$                                &  $2.6_{-0.4}^{+0.5}$   & $2.1_{-0.2}^{+0.4}$    & $2.1_{-0.2}^{+0.2}$\\

&   Mg  & $1.3_{-0.2}^{+0.2}$                                                   &  $1.3_{-0.1}^{+0.2}$      &  $1.4_{-0.2}^{+0.1}$      & $1.2_{-0.2}^{+0.2}$ \\

&  Si  & $1.5_{-0.2}^{+0.2}$                                                    &  $1.2_{-0.1}^{+0.3}$      & $1.6_{-0.2}^{+0.2}$       & $1.6_{-0.3}^{+0.2}$  \\

&  S  & $1.6_{-0.2}^{+0.2}$                                                     &  $1.8_{-0.2}^{+0.4}$      &  $1.9_{-0.2}^{+0.3}$      & $1.2_{-0.1}^{+0.2}$\\

&  Fe  & $1.6_{-0.5}^{+0.6}$                                                      &  $1.5_{-0.4}^{+0.4}$      &  $1.4_{-0.2}^{+0.3}$      & $1.4_{-0.2}^{+0.1}$\\

&   $\tau$=$n_{\rm e}t$ ($10^{10}$ cm$^{-3}$ s)&  $4.1_{-0.5}^{+0.6}$   &$3.3_{-0.6}^{+0.4}$        &  $5.2_{-0.4}^{+0.4}$      & $4.3_{-0.6}^{+0.4}$ \\

&   Norm (10$^{-3}$ cm$^{-5}$) &  $4.8_{-0.7}^{+0.6}$   & $0.7_{-0.1}^{+0.2}$    & $2.4_{-0.1}^{+0.1}$    & $2.4_{-0.2}^{+0.5}$\\
\hline
& reduced-$\chi^{2}$ (dof) & 1.19 (1541)                                           &  1.04 (309)               & 1.09 (793)                &  1.13 (1056) \\
 \hline
\end{tabular}
\begin{tablenotes}
\item {\bf Notes.} The normalization of the VNEI, norm=$10^{-14}$$\int n_{\rm e} n_{\rm H} dV$/($4\pi d^{2}$) (cm$^{-5}$), where $n_{\rm e}$, $n_{\rm H}$, $V$ and $d$ are the electron and hydrogen densities (cm$^{-3}$), emitting volume (cm$^{3}$) and distance to the source (cm), respectively. Abundances are given relative to the solar values of \citet{Wi00}.   
\end{tablenotes}
\end{threeparttable}
\end{minipage}
\end{table*}

\begin{figure*}
\centering \vspace*{1pt}
\includegraphics[width=0.75\textwidth]{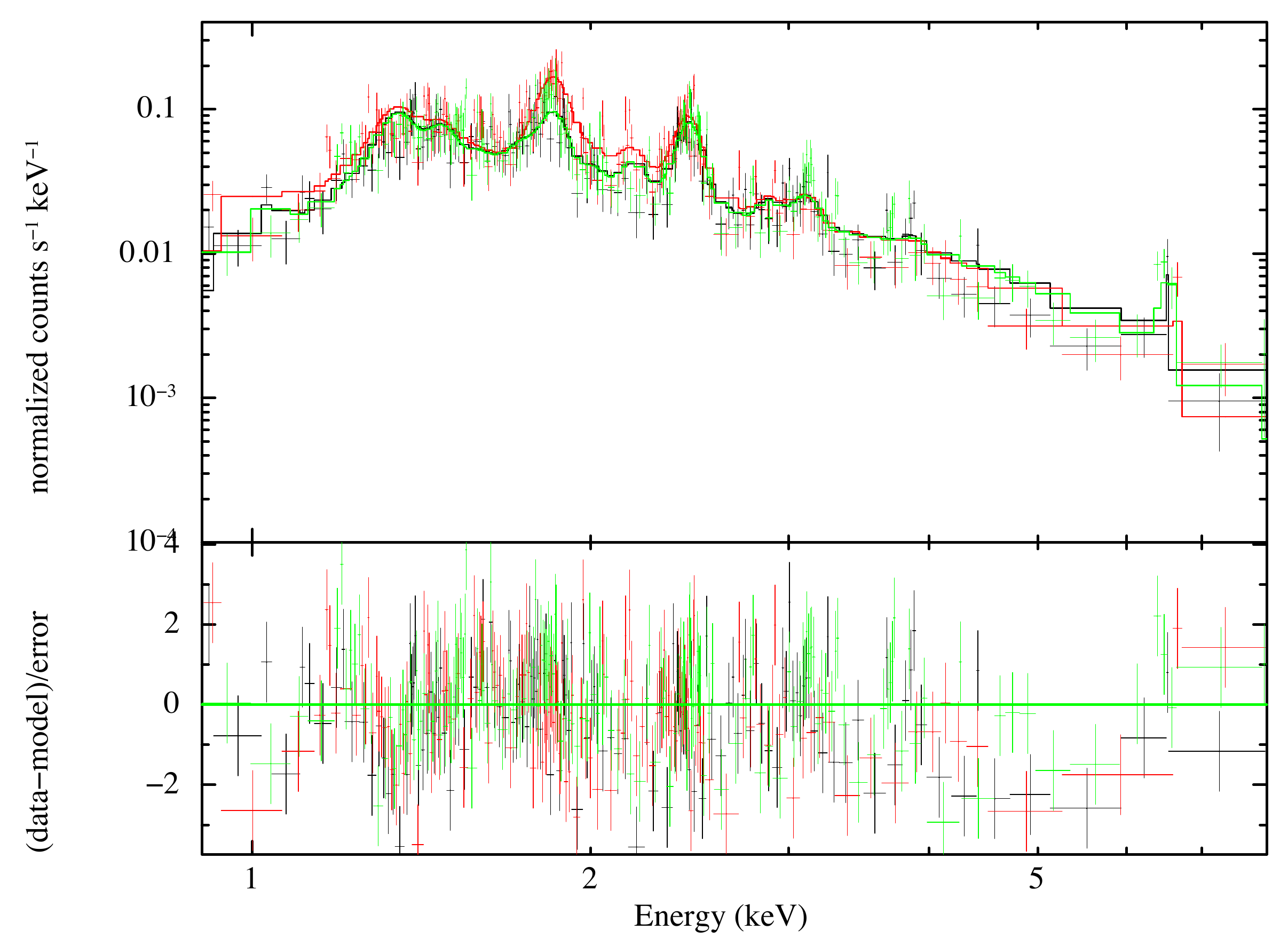}
\caption{{\it Suzaku} XIS0, XIS1 and XIS3 (in black, red and green, respectively) spectra of Kes 69 (whole SNR) as fitted with a TBABS$\times$VNEI model in the 0.9$-$8.0 keV energy band. Solid lines indicate the best-fitting VNEI model. Lower panel shows the $\chi^{2}$ fit residuals. The parameters of the best-fitting spectra are given in Table 1.}
\label{figure_2}
\end{figure*}

\subsection{Gamma-ray Analysis}
We used \texttt{fermipy}\footnote{http://fermipy.readthedocs.io/en/latest/index.html} analysis toolkit to apply the maximum likelihood fitting method \citep{Ma96} on spatially and spectrally binned data for 0.2 $-$ 300 GeV by taking P8R2$_{-}$SOURCE$_{-}\!\!$V6 as the instrument response function. The analysis yields the test statistics (TS) value, which is the square root of the detection significance. A large TS value shows that the maximum likelihood value for a model without an additional source (the null hypothesis) is not valid. 

The background model of the analysis region consists of the diffuse background sources and all the point-like and extended sources from the 3rd {\it Fermi}-LAT Source Catalog \citep{Ac15} located within 15$^{\circ}$ $\times$  15$^{\circ}$ region centred on the ROI centre. All parameters of the diffuse Galactic emission (\emph{gll$_{-}$iem$_{-}$v6.fits}) and the isotropic component (\emph{iso$_{-}$P8R2$_{-}$SOURCE$_{-}\!\!$V6$_{-}\!$v06.txt}) were freed. The normalization parameters of all sources within 3$^{\circ}$ are set free. In addition, we freed all sources with TS $>$ 10 and fixed all sources with TS $<$ 10.

\section{Results and Discussion}
\subsection{X-ray Properties of Kes 69}
The X-ray spectra of the SNR were previously studied by \citet{Yu03} with {\it ROSAT} and {\it ASCA}, by \citet{Bo12} and by \citet{Se13} with {\it XMM-Newton}. Although their spectral fits indicate that the X-ray-emitting gas is characterized by a CIE model, we find that the plasma is well represented by a NEI plasma model. Our estimated absorption is slightly higher than {\it ROSAT} and {\it XMM-Newton} results. The CIE plasma temperatures found are $\sim$1.6 keV \citep{Yu03}, $\sim$0.8 keV \citep{Bo12} and $\sim$0.62 keV \citep{Se13}, which are significantly lower than our result. These inconsistencies may be related to the detection of Fe-K line in the {\it Suzaku} spectra, which produces significantly higher $kT_{\rm e}$ compared with previous studies. Also, in our analysis we use different background regions, different background estimation method, the different spectral capabilities of the X-ray satellites, the elemental abundances by \citet{Wi00} and the recently updated AtomDB database.

The X-ray spectra of the annular regions surrounding the X-ray centre of Kes 69 show no strong evidence of spectral variability. A characteristic feature of MM SNRs is a flat electron temperature profile \citep{Rh98}. Kes 69 shows a flat temperature of $\sim$2.3 keV across its centre, which is consistent with MM SNRs. The ionization time-scales for all regions indicate that the X-ray emitting plasma across the whole remnant is in an NEI condition. We find enhanced abundances of Mg, Si, S and Fe in all regions indicating that the X-ray emitting plasma has an ejecta origin. However, this result is not conclusive when the error bars are taken into account (see Table 1).

The density of the X-ray emitting gas was estimated from the normalization, norm=$n_{\rm e}n_{\rm H}
V$/($4\pi d^{2}$$10^{14}$). Assuming that the emitting region to be a
sphere of radius 6.1 arcmin, the SNR is at a distance of
5.2 kpc and $n_{\rm e}=1.2n_{\rm H}$, we estimated the emission
volume to be $V\sim 9.6\times10^{58}fd_{5.2}^{3}$ ${\rm cm^{3}}$, where $f$ is the volume filling factor and $d_{5.2}$ is the distance scaled to 5.2 kpc. Consequently, we found an ambient gas density of $\sim$0.14$f^{-1/2}d_{5.2}^{-1/2}$
${\rm cm}^{-3}$ and age of $\sim$$9.2\times10^{3}f^{1/2}d_{5.2}^{1/2}$ yr. Our estimated age of the remnant is comparable to the derived by \citet{Bo12} and lower than that derived by \citet{Se13} in their {\it XMM-Newton} analysis. We then derived the total X-ray-emitting mass, $M_{\rm X}$ $\sim 15.8 f^{1/2}d_{5.2}^{5/2}{M\sun}$ using $M_{\rm X}$=1.4$m_{\rm H}n_{\rm e}V$. This result indicated that the X-ray-emitting plasma is dominated by ejecta material of a core-collapse (CC) supernova (SN), because the X-ray-emitting mass is much larger than the Chandrasekhar mass.

Assuming the \citet{Se59} model, we calculated the SNR radius of $R$ $\sim$ 18.6 pc and the swept-up interstellar medium (ISM) mass of $M_{\rm sw}$ $\sim$ $56M_{\sun}$ using the initial ISM density $n_{0}$=0.14 cm$^{-3}$, SNR age $t=9200$ yr and explosion energy $E=10^{51}$ erg. The swept-up ISM mass of Kes 69 is reasonably explained if the X-ray-emitting plasma is dominated by the shocked ambient medium. We note that a large swept-up ISM mass may be expected for MM SNRs, which appear to be interacting with MCs.

We also calculated the total thermal energy of $\sim$$10^{51}$ erg by assuming $M$=56$M_{\sun}$ and ion temperature $T_{\rm i}$=2.5 keV. We therefore cannot rule out the ISM origin if $T_{\rm e}$=$T_{\rm i}$. However, the relaxation time-scale due to the Coulomb interaction between $T_{\rm i}$ and $T_{\rm e}$ is about $10^{13}$ sec ($n$/0.1 cm$^{-3}$)$^{-1}$, which is much larger than expected age of the SNR. So, we may expect $T_{\rm i}$ $>>$ $T_{\rm e}$ and the ISM model requires explosion energies much larger than $10^{51}$ erg.Ê In this case, the X-ray emitting plasma would be more like ejecta in origin.Ê

\subsection{Gamma-ray Results}
In the energy range of 0.2 - 300 GeV no excess gamma-ray emission was found from the direction of Kes 69. The upper limit at 95 per cent confidence level (CL) on the flux and energy flux were found to be 3.1 $\times$ 10$^{-6}$ MeV cm$^{-2}$ s$^{-1}$ and 2.1 $\times$ 10$^{-9}$ cm$^{-2}$ s$^{-1}$, respectively. The flux upper limit is comparable to the upper limit given in the 1st {\it Fermi}-LAT Supernova Remnant Catalog \citep{Ac16}, which is 1.8 $\times$ 10$^{-9}$ cm$^{-2}$ s$^{-1}$ for a spectral index of 2.5 and 95 per cent upper limit. Assuming the distance of 5.5 kpc to the SNR, the upper limit of the gamma-ray luminosity is 1.8 $\times$ 10$^{34}$ erg s$^{-1}$, which is slightly smaller than the reported gamma-ray luminosities for interacting SNRs \citep{Ac16}.

Kes 69 is an MM SNR in Sedov phase and shows signs of interaction with MC thorough the detection of OH (1720 MHz) masers. However, due to narrow physical conditions needed to excite the OH (1720 MHz) masers, the lack of them would not give any information about hadronic gamma rays being emitted from this SNR or not \citep{Fr11}. Besides, although the cosmic rays may be responsible for the local enhancement in SNRs, which is sufficient to produce OH abundance in the post shock gas, the ionization rate from the interior X-ray emission is comparable to the one from cosmic rays. The dominant emission mechanism might depend on the gas location with respect to the interior X-ray emitting plasma and cosmic ray acceleration site \citep{He09b}.

\citet{He08} reported extended OH (1720 MHz) maser emission from Kes 69 using GBT data, which was previously not detected by VLA (summarized in Table 3 of \citet{He08}). They reported a single compact maser emission at +69 km~s$^{-1}$, which was also detected during the VLA observations \citep{Gr97}. However, faint OH (1720 MHz) emission is present at +85 km~s$^{-1}$ that appears across the southern ridge of the SNR, which has broader line widths and a velocity gradient. Since both the extended maser emission and molecular material are found at around +85 km~s$^{-1}$, it raises suspicion about the Kes 69 origin of the maser found at +69 km~s$^{-1}$. If +69 km~s$^{-1}$ maser is not associated with Kes 69, then the OH (1720 MHz) maser emission arising from Kes 69 is only extended in nature. These extended masers are about 20 times lower in brightness than compact masers and they may be produced by the interior X-ray emission of the SNR. \citet{Zh09} showed that a molecular arc is present at 77-86 km s$^{-1}$ at the southeast part of the shell of Kes 69, which overlaps with the 1.4 GHz radio continuum and mid-infrared (IR) observations at the same region of the shell. They also found HCO$^{+}$ emission at 85 km s$^{-1}$ on the shell. Both the molecular arc and the HCO$^{+}$ emission at $\sim$85 km s$^{-1}$ were reported to be consistent with the presence of the extended 1720 MHz OH emission along the southeastern boundary of Kes 69.

According to \citet{Zh09}, the multi-wavelength emissions along the southeastern shell of Kes 69 were caused by the impact of the SNR shock on a dense, clumpy patch of molecular gas, which pre-existed and is possibly the cooled and clumpy left-over debris of the interstellar molecular gas swept up by the progenitors stellar wind. Therefore, a possible explanation of the lack of GeV gamma-ray emission from Kes 69 is that the extended maser emission is produced by the interior X-ray emitting plasma rather than by cosmic rays accelerated at regions away from the molecular gas clumps.

\subsection{Kes 69: A Maser Emitting Mixed-morphology SNR}

Kes 69 is a middle-aged SNR, its centre-filling X-ray emission surrounded by a shell-like radio structure suggests that Kes 69 belongs to the category of MM SNRs. There are two main models to explain the centrally filled X-ray emission in the MM SNRs. One is the evaporating cloudlet model (e.g., \citealt{Wh91}) and the other is the thermal conduction model (e.g., \citealt{Co99}). But, the true origin of the centre-filled thermal X-ray emission from middle-aged MM SNRs is still unknown. It is unclear what leads to their centre-filled morphology and whether these SNRs are dominated by the ejecta or by the shocked ISM. The origin of emission from ejecta-dominated MM SNRs also remains unknown. Our XIS analysis shows that Kes 69 is likely to be of ejecta origin.

Recently, \citet{Sl17} performed multi-dimensional hydrodynamics simulations to discuss the origin of centrally peaked X-ray profile. They do not include any piston ejecta but put thermal energy in their simulations. Then the central diffuse gas expands due to large thermal pressure, pushing ISM to form the shock. According to \citet{Sl17} the origin of the hot gas is the ISM and not the ejecta. In addition, \citet{Sl17} suggest that the SNR's centre-filled morphology is due to the expansion into a cloudy medium and ejecta is not a dominant factor in the formation of this morphology. As mentioned in Section 4.1, X-ray properties of Kes 69 show that the detected hot gas in Kes 69 is likely ejecta in origin rather than ISM.

Here, we discuss Kes 69 in comparison with other MM SNRs known to be interacting with MCs. We list the SNRs identified as interacting MM SNRs type and summarize their properties, such as detection of RP and gamma-ray in Table 2. As seen in Table 2, all of the SNRs that have detected RP emission, show also gamma-ray emission except for G346.6$-$0.2. Other interacting MM SNRs, like W51C and G298.6$-$0.0 in Table 2, show gamma-ray emission, but no RP emission. There has been no published analysis results about RP in Sgr A East and Kes 41, but they present gamma-ray emission. However, Kes 69 is the only candidate in the list for which no gamma-rays nor RP emission is detected.

\begin{table*}
 \begin{minipage}{240mm}
  \caption{RP and gamma-ray detection in interacting MM SNRs.} 
 \begin{threeparttable}
\begin{tabular}{@{}p{1.5cm}p{1.4cm}p{1.6cm}p{1.4cm}p{1.5cm}p{1.3cm}p{1.4cm}p{1.0cm}p{2.5cm}@{}}
  \hline\hline
SNR       & Other                 & ${\gamma}$-ray           &  RP        &  MC &  $kT_{\rm init}$ & $kT_{\rm e}$ & Age$^{\rm d}$    &      References \\ 
       &          name        &     detection     &   detection    &  interaction$^{\rm a}$  & (keV)& (keV)& (kyr)    &      \\
\hline

G0.0+0.0 & Sgr A East      &         GeV/TeV           &      ?       &      Y     &    -     &   $1.21_{-0.03}^{+0.02}$, $6.0_{-0.5}^{+0.4}$    & 1.2$-$10 &  [1, 2, 3, 4]   \\   

G6.4$-$0.1& W28           &    GeV/TeV              &    Y         &        Y  &3 (fixed)& $0.40_{-0.03}^{+0.02}$  & 33$-$36  &    [5, 6, 7]       \\

G21.8$-$0.6$\dagger$& Kes 69         &     N               &     N         &        Y  &  $0.57_{-0.02}^{+0.03}$ &  $2.57_{-0.34}^{+0.26}$ & 5$-$25 &    This work      \\

G31.9+0.0 & 3C 391          &       GeV              &    Y         &        Y    &$1.8_{-0.6}^{+1.6}$& 0.495$\pm$0.015 & 3.7$-$4.4 &    [8, 9]       \\

G34.7$-$0.4$\dagger$ & W44            &     GeV             &    Y      &        Y   &$1.07_{-0.06}^{+0.08}$& 0.48$\pm$0.02   & 20$-$28 &    [10, 11]       \\ 

G49.2$-$0.7& W51C         &          GeV/TeV          &    N        &       Y  &  -  & $0.69_{-0.05}^{+0.06}$  & 18$-$30  &   [12, 13, 14]       \\  

G89.0+4.7$\dagger$ &HB21           &          GeV           &      Y       &       Y$^{\rm b}$  &  $0.58_{-0.07}^{+0.09}$ & $0.17_{-0.02}^{+0.01}$ &  4.8$-$15 &  [15, 16]     \\ 

G189.1+3.0$\dagger$& IC 443        &  GeV/TeV		 &       Y      &      Y   &  10 (fixed) & 0.65$\pm$0.04& 3$-$30 &   [17, 18, 19, 20]        \\  

G298.6$-$0.0 &           &          GeV           &      N       &     Y$^{\rm c}$  &  - &  $0.78_{-0.08}^{+0.09}$ &  - &  [21]        \\

G337.8$-$0.1 &Kes 41     &      GeV		 &      ?       &      Y    &  - &  $1.80_{-0.47}^{+0.82}$ &  12$-$16 &  [22, 23]     \\  

G346.6$-$0.2&		   &     	N	 &       Y      &       Y  &  5 (fixed)& $0.30_{-0.01}^{+0.03}$   &   4.2$-$16 & [24, 25]        \\

G348.5+0.1 &CTB 37A      &  GeV	 	 &   	  Y     &      Y  & 5 (fixed)& $0.49_{-0.06}^{+0.09}$ & 10$-$30  & [26, 27]        \\  

G359.1$-$0.5$\dagger$ &              &       GeV/TeV              &         Y    &        Y & $0.77_{-0.08}^{+0.09}$& 0.29$\pm$0.02  & $\geq$10 &  [28, 29, 30]       \\  
 \hline
\hline
\end{tabular}
\begin{tablenotes}
\item {\bf Notes.} $^{\rm a}$OH detection from \citet{Ji10}. $^{\rm b}$Interaction shown by CO MA \& LB, CO ratio, H$_{2}$, NIR. $^{\rm c}$Possible detection of IR emission  \citep{Re06}. $^{\rm d}$SNR catalog in \citet{Fe12}. $\dagger$ Ejecta-dominated. 

\item {\bf References.} [1] \citet{Ah06}; [2] \citet {Ac16}; [3] \citet{Aj17}; [4] \citet{Ko07b}; [5] \citet {Ah08a}; [6] \citet{Ab10c}; [7] \citet{Sa12}; [8] \citet {Er14}; [9] \citet{Sa14}; [10] \citet{Ab10b}; [11] \citet{Uc12}; [12] \citet {Ab09}; [13] \citet{Al12}; [14] \citet{Ha13}; [15] \citet{Pi13}; [16] \citet{Su18};  [17] \citet{Ab10a}; [18] \citet{Alb07}; [19] \citet{Ya09}; [20] \citet{Oh14}; [21] \citet {Ba16}; [22] \citet{Li15}; [23] \citet{Zh15};  [24] \citet{Er12}; [25] \citet{Ya13}; [26] \citet{Ca10}; [27] \citet{YaS14}; [28] \citet{Ah08b}; [29] \citet{Hu11}; [30] \citet{Oh11}. 
\end{tablenotes}
\end{threeparttable}
\end{minipage}
\end{table*}

\begin{table*}
\caption{Best-fitting abundances of Mg, S and Fe relative to Si with the expected relative abundances for CC and Type Ia SNe models.}
 \begin{minipage}{140mm}
\begin{center}
\begin{threeparttable}
\renewcommand{\arraystretch}{1.5}
 \begin{tabular}{@{}cccccccccccc@{}}
  \hline
    \hline
    &&\multicolumn{4}{c}{Type Ia Models}&\multicolumn{6}{c}{CC Models$^\ast$}\\
    \cline{3-6}
     \cline{7-12}
      Ratio& Kes 69 & W7$\dagger$
      & WDD2$\dagger$& PDDe$\ddagger$
      & DDTe$\ddagger$ &11$M_{\sun}$ &12$M_{\sun}$ &15$M_{\sun}$ & 20$M_{\sun}$  &  25$M_{\sun}$  &30$M_{\sun}$ \\
\hline

Mg/Si     & $0.81_{-0.18}^{+0.20}$     &  0.06     & 0.02     & 0.0017$^\star$  & 0.025$^\star$     & 0.57     & 0.12   & 0.70    & 0.16    & 0.45    & 1.79    \\

S/Si      & $1.02_{-0.18}^{+0.19}$     &  1.07     & 1.17     & 1.5     & 1.4     & 0.87     & 1.53   & 0.62    & 1.28    & 0.96    & 0.24    \\
 
Fe/Si     & $1.06_{-0.35}^{+0.39}$     &  1.56     & 0.85     & 0.89    & 0.91    & 1.37     & 0.23   & 0.70    & 0.88    & 0.13    &  0.15    \\
  \hline
\end{tabular}
\begin{tablenotes}
\item {\bf Notes.} $\dagger$ \citet{No97}. $\ddagger$ \citet{Ba03}. $^\ast$\citet {Wo95}. $^\star$\citet {Ra06}.
\end{tablenotes}
\end{threeparttable}
\end{center}
\end{minipage}
\end{table*}

\subsection{The explosion origin of Kes 69}
As mentioned before, Kes 69 is an ME SNR and {\it Spitzer} detection of spectral lines associated with shocked H$_2$, which are strong evidences of CC origin. We consider other methods to estimate the progenitor for this SNR. There are several methods to determine the origin of SNRs; (i) X-ray morphology: Using high-resolution {\it Chandra} images of SNRs, \citet{Lo09, Lo11} found that the CC SNRs are more spatially asymmetric than the Type Ia SNRs; (ii) Fe-K line: \citet{Ya14} concluded that Fe-K centroid energies for Type Ia SNRs are lower ($<$6.55 keV) than those of CC SNRs; (iii) Abundance of ejecta; (iv) Presence of compact object or pulsar wind nebulae (PWN); (v) Environment. 
 
To determine the Fe-K centroid energy, we fitted the 5.0$-$8.0 keV spectrum with an absorbed power-law plus a Gaussian model. We note that the Gaussian line width parameter was fixed to zero. For comparison with the results of \citet{Ya14}, we estimated the centroid energy of Fe-K$\alpha$ as $6.47_{-0.03}^{+0.03}$, $6.50_{-0.06}^{+0.05}$, $6.49_{-0.04}^{+0.04}$ and $6.45_{-0.05}^{+0.05}$ keV for whole, Regions 1, 2 and 3, respectively, which agree on a Type Ia origin.

The GRXE also emits K-shell lines from neutral iron (at 6.4 keV), helium-like iron (at 6.7 keV) and hydrogen-like iron (at 7.0 keV) (e.g. \citealt{Uc13}). Therefore, we compare the flux of Fe-K line from Kes 69 and GRXE. The Gaussian line at $\sim$6.5 keV of the remnant has a surface brightness of $\sim$0.8$\times$10$^{-7}$ photons cm$^{-2}$ s$^{-1}$ arcmin$^{-2}$, while that of the background is very faint (less than $\sim$1 per cent of the $\sim$6.5 keV line flux detected in the remnant). This result indicates that 6.5 keV Fe-K line is not a contamination of the Fe K-shell line in the GRXE and the detection of the $\sim$6.5 keV line feature is robust.

Table 3 presents the best-fitting abundances of the whole region relative to Si, the abundance ratios expected for CC SN with different progenitor masses CC \citep{Wo95} and Type Ia models \citep{No97, Ba03}. Mg/Si ratios are taken from \citet{Ra06}, since they were not given in \citet{Ba03}. As seen from Table 3, we find that the CC model with 11$M_{\sun}$ is feasible.

We also plotted the element masses in the X-ray-emitting gas, the Type Ia models of \citet{No97} and \citet{Ba03} and the CC nucleosynthesis models with various progenitor masses of \citet{Wo95, No06, Th96, Su16} and \citet{Fr18} for comparison in Figure 3 and 4, similar to \citet{Zh16}. In some CC models (e.g., \citealt{Fr18}), the yields are strongly dependent on the SN explosion energy. To see how the abundance depends on the progenitor mass and the explosion energy, we plotted the yields for different SN explosion energies (15, 20 and 25$M_{\sun}$)  as given by \citet{Fr18}, which is shown in Figure 4. We see that none of the 25$M_{\sun}$ models can explain the observed the elemental mass for this SNR. As shown in Figure 3 and 4, we found that the estimated values of $M_{\rm Mg}$ $\sim$ 0.01, $M_{\rm Si}$ $\sim$ 0.016, $M_{\rm S}$ $\sim$ 0.009 and $M_{\rm Fe}$ $\sim$ 0.0018 $M_{\sun}$ are in good agreement with the progenitor mass of 11$M_{\sun}$ \citep{Wo95} and between 9 and 12$M_{\sun}$ \citep{Su16} of the CC models.

The environment of Kes 69 is suggestive of a massive progenitor star: it is interacting with MCs and the detection of IR emission from shocked molecular gas. Also its X-ray morphology has highly asymmetric nature and it is located in the Galactic plane. All these environmental and morphological typing methods support that Kes 69 has a CC origin. We note that there are no features for central compact object or a PWN in the remnant. 

In summary, based on the total ejecta mass (as discussed in Section 4.1) and the result of our comparison between the element masses in the X-ray-emitting plasma and the predicted SN yields (as seen in Figure 3), we conclude that the origin of Kes 69 is a CC SN.

The centroid energy of Fe-K$\alpha$ line ($\sim$6.5 keV) corresponds to a charge number of about +18 (see Figure 1, left; \citealt{Ya14}), which is consistent with $n_{\rm e}t$ $\sim$ $4\times10^{10}$ cm$^{-3}$ s (see Figure 5 of \citealt{Ha15}). The low $n_{\rm e}t$ value and low centroid energy of Fe-K$\alpha$ suggest that the CC SN exploded in the low density region.

 \begin{figure*}
\includegraphics[width=0.49\textwidth]{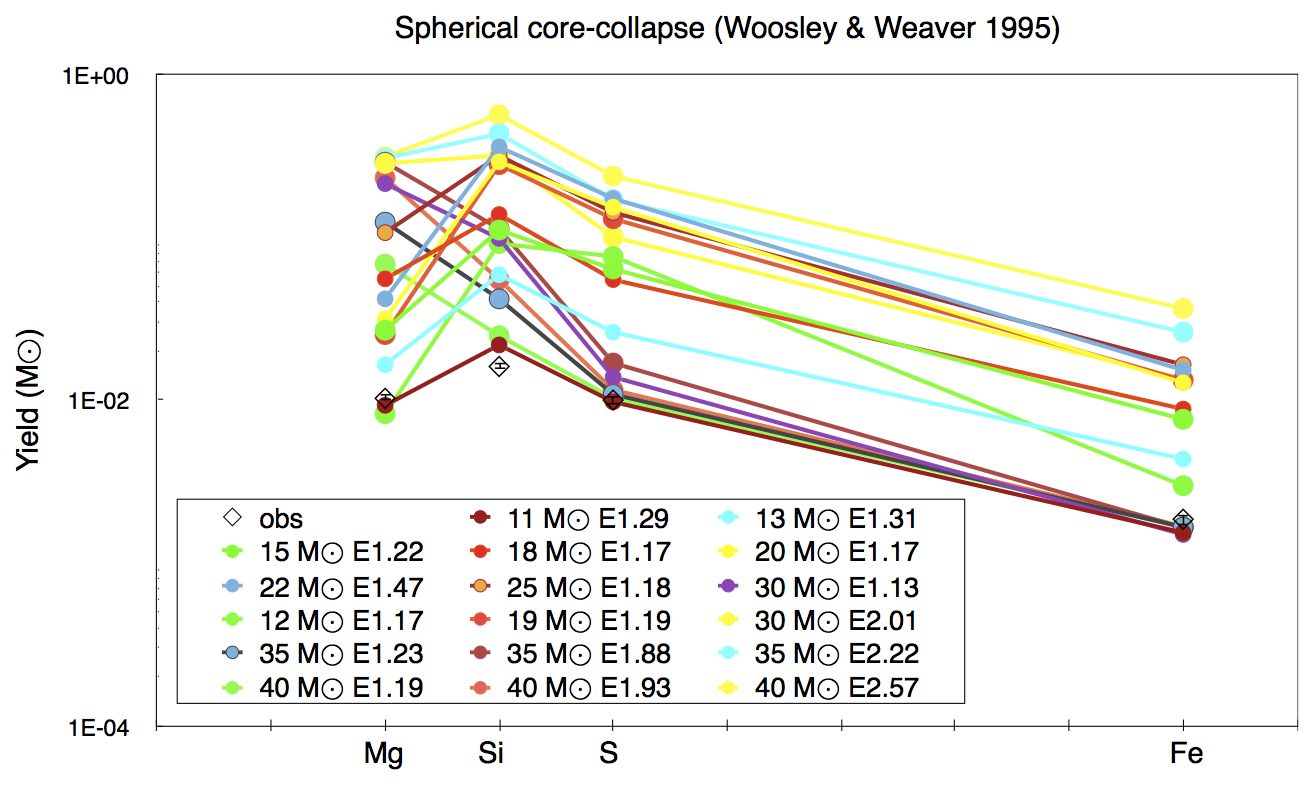}
\includegraphics[width=0.49\textwidth]{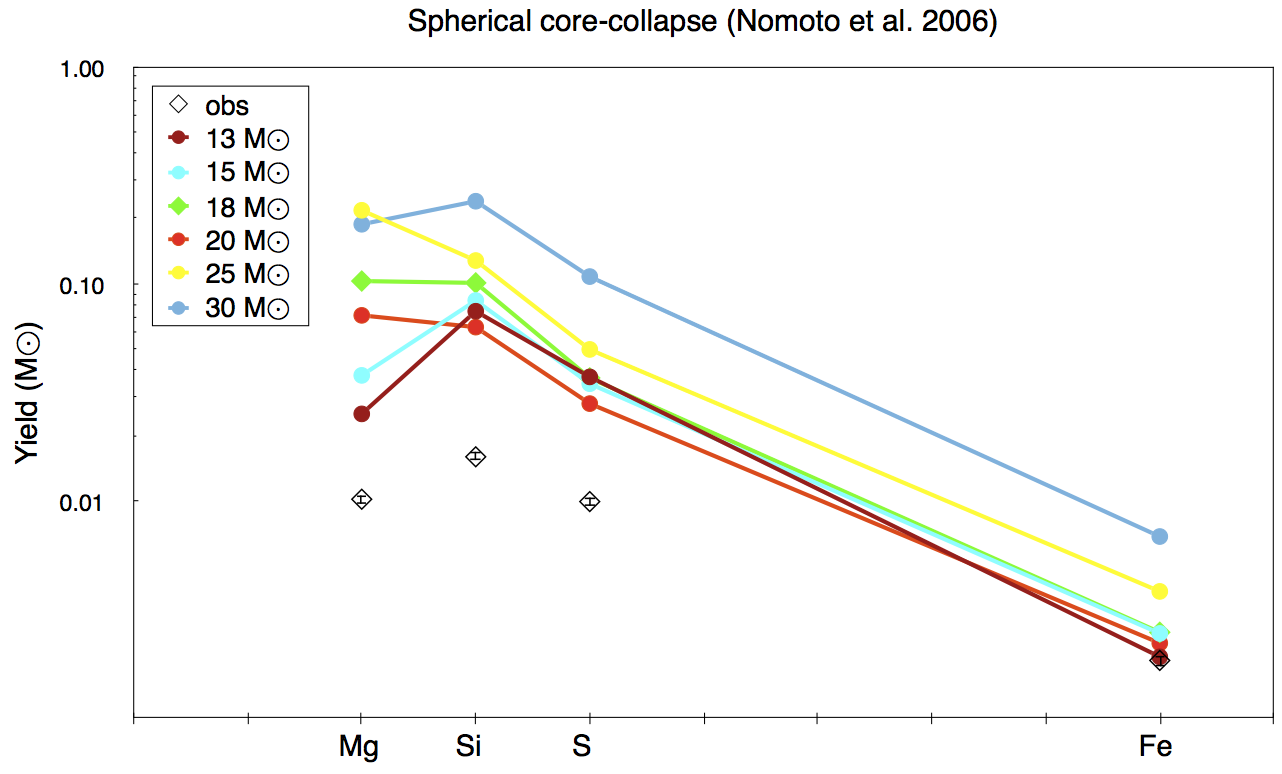}
\includegraphics[width=0.49\textwidth]{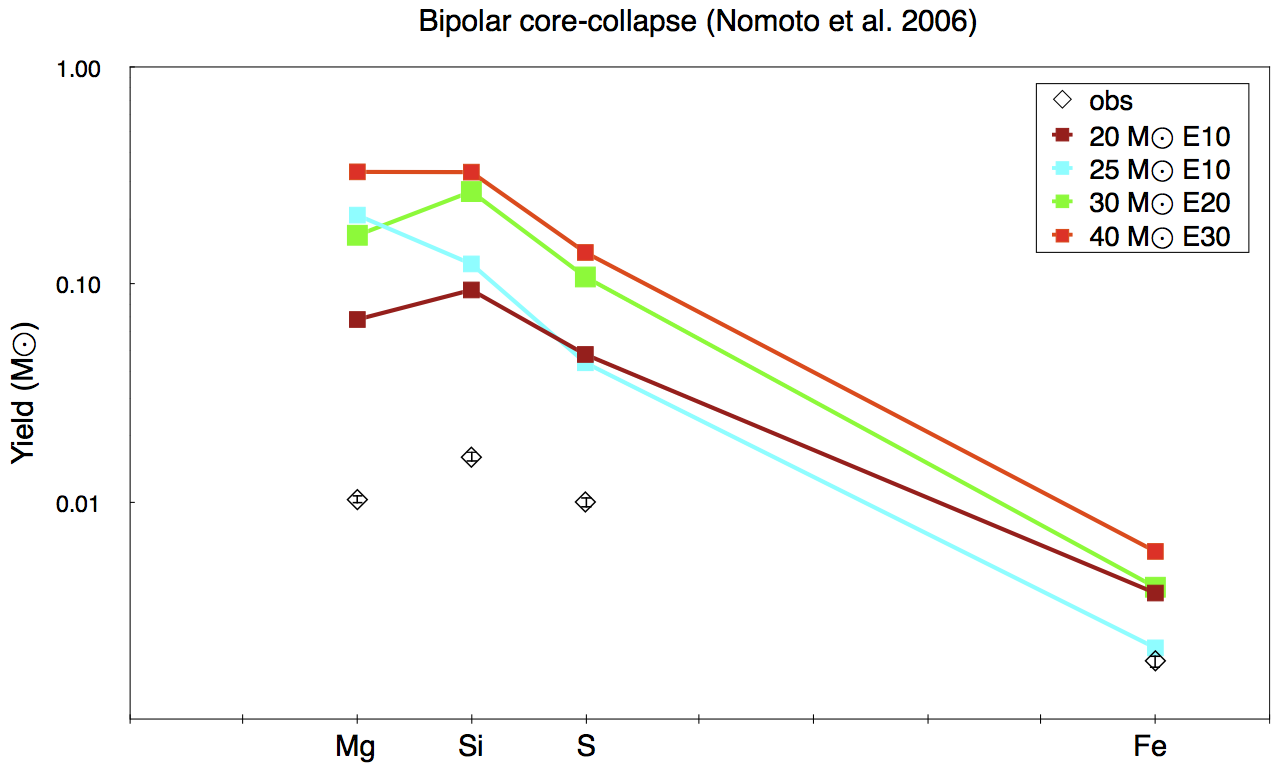}
\includegraphics[width=0.49\textwidth]{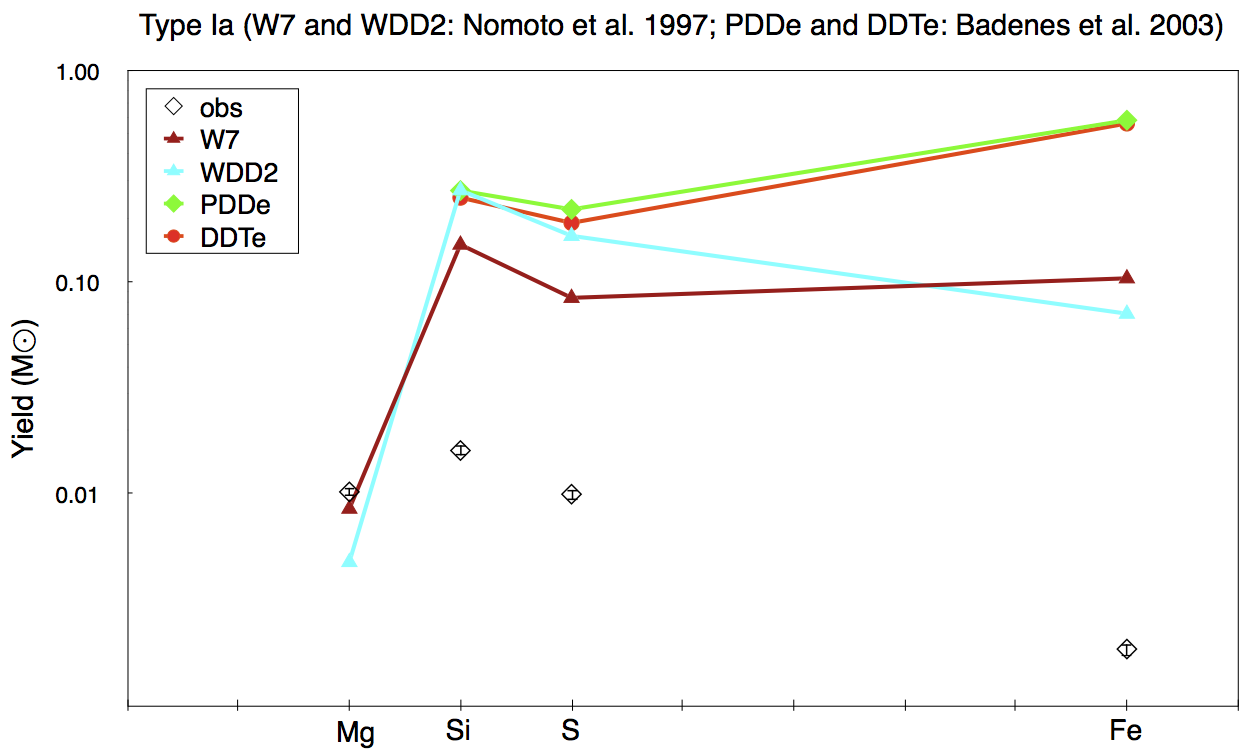}
\includegraphics[width=0.49\textwidth]{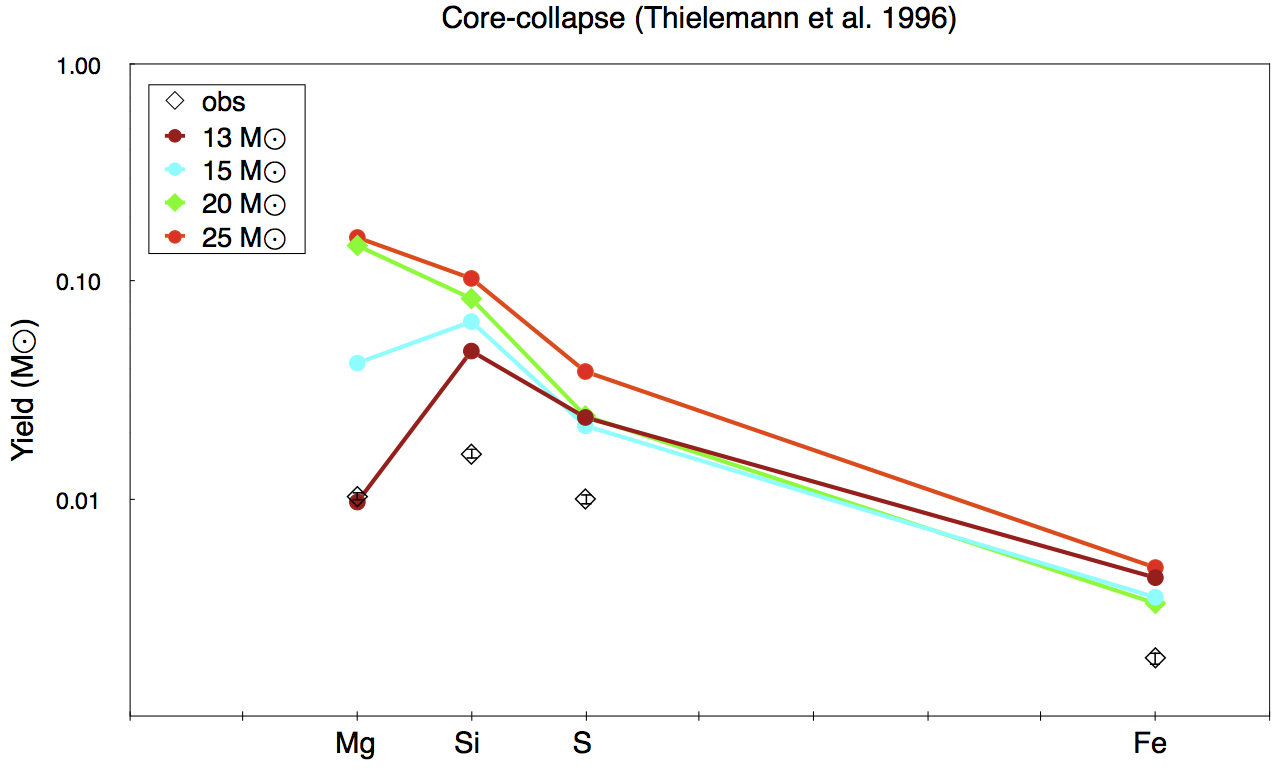}
\includegraphics[width=0.49\textwidth]{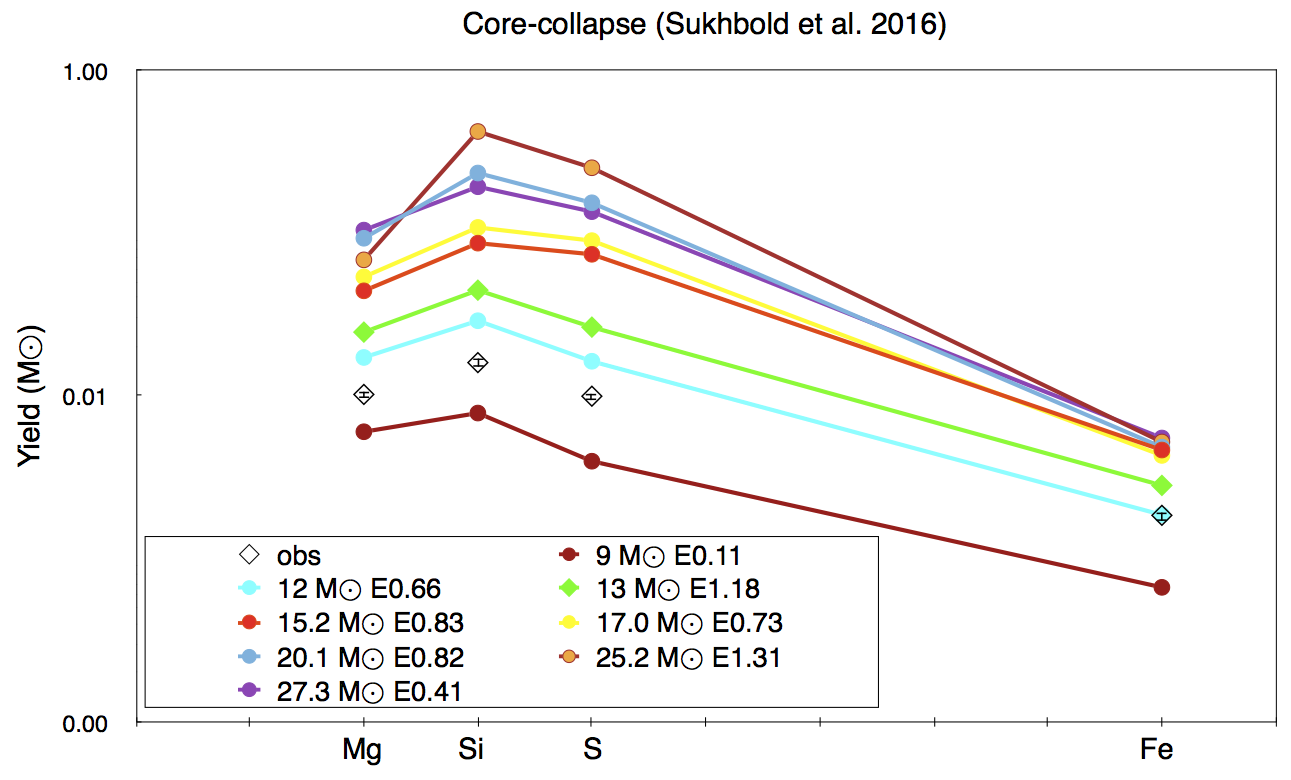}
\caption{The masses of the Mg, Si, S and Fe of the X-ray emitting gas (diamonds with error bars) compared with the predictions of SN nucleosynthesis models. The CC SN models yields from \citet{Wo95, No06, Th96, Su16} and \citet{Fr18}. The Type Ia models (W7, WDD2, PDDe and DDTe) yields from \citet{No97} and \citet{Ba03}. $E$ is the explosion energy (e.g., in the left top panel $E1.29$ stands for 1.29$\times10^{51}$ erg).}
\label{figure_3}
\end{figure*}

 \begin{figure*}
\includegraphics[width=0.55\textwidth]{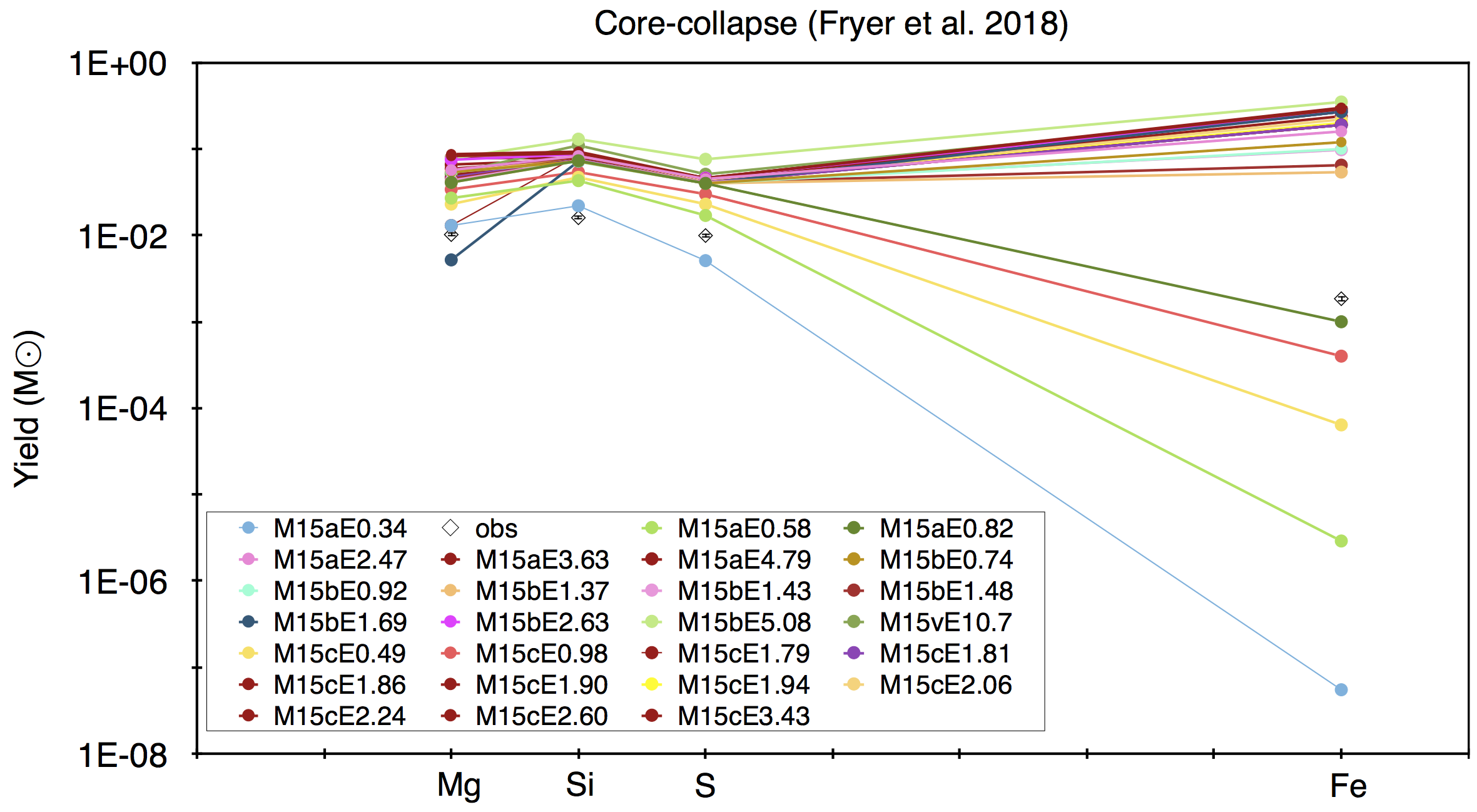}
\includegraphics[width=0.55\textwidth]{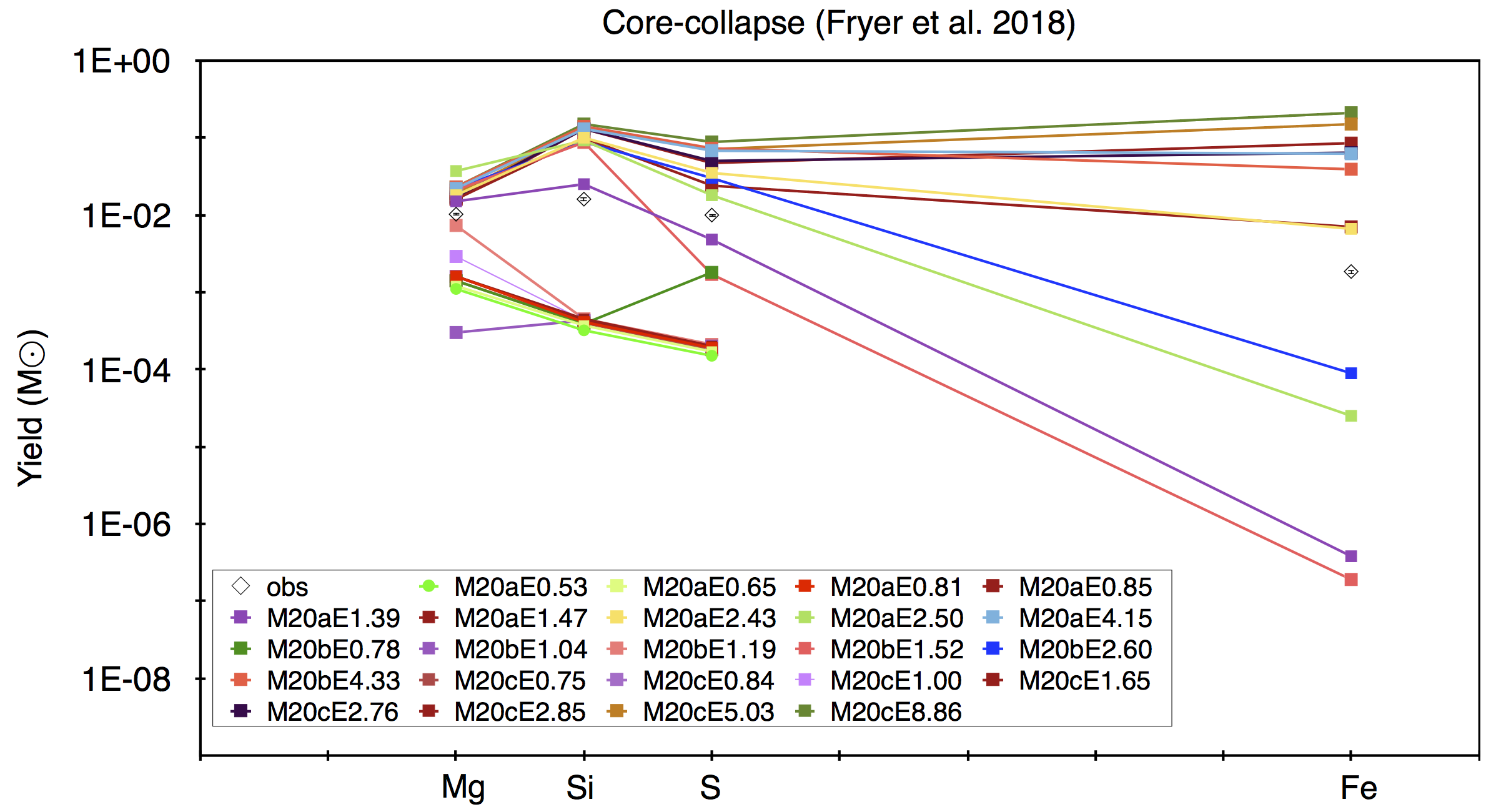}
\includegraphics[width=0.55\textwidth]{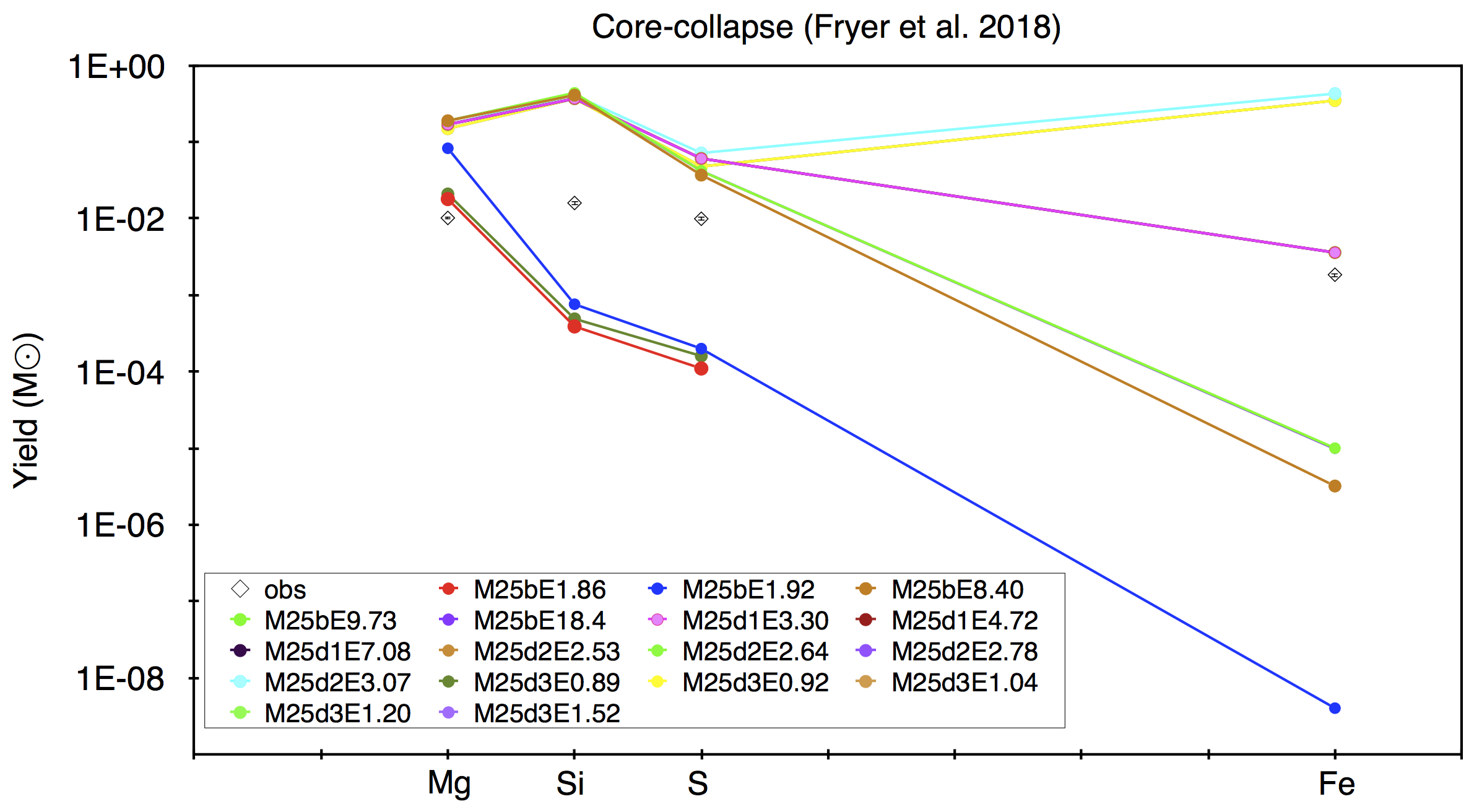}
\caption{Continued from Figure 3.}
\end{figure*}

\section{Conclusions}

In this paper, we analyze the archival {\it Suzaku} and {\it Fermi}-LAT data of Kes 69 and report on the spectral properties of the X-ray emission of this remnant. Our main conclusions are summarized below:
 
\begin{enumerate}

\item The centre-filled X-ray emission and the radio shell-like morphology suggest that Kes 69 is a member of the MM SNR class, as noted before by \citet{Yu03}.

\item We confirmed the thermal origin of the central plasma and detected slightly enhanced abundances of Mg, Si, S and Fe, which suggest the possible presence of SN ejecta in Kes 69. Thus, we concluded that Kes 69 is an ejecta-dominated MM SNR candidate. Future X-ray observations with the superb high angular resolution satellites like {\it Chandra} will allow us to study the X-ray structure of the ejecta and the surrounding medium.

\item The spectra can be well described by a single NEI model with a temperature of $\sim$$2.5$ keV. The ionization time-scale is $\sim$4.1$\times10^{10}$ cm$^{-3}$ s, which implies that the plasma is far from ionization equilibrium, still ionizing. We found no spectral variation across the SNR. We detect the Fe-K$\alpha$ emission line at $\sim$6.5 keV in the spectra.

\item We searched for RP in Kes 69 and found that the initial plasma temperature much smaller than the current plasma temperature. Therefore, we concluded that the X-ray emitting plasma of Kes 69 is in the NEI state, not overionized.

\item We investigated the origin of SN type and concluded that the origin of Kes 69 is a CC SN and its progenitor mass is most likely between 9 and 12$M_{\sun}$ (see Figure 3).

\item We found no significant gamma-ray emission detected from Kes 69. The lack of gamma-ray emission from this interacting MM SNR can be explained by the dominant emission mechanism depending on the location of the gas with respect to the X-ray emitting plasma and the cosmic ray acceleration site. Assuming that the only associated OH (1720 MHz) maser emission is the extended maser emission at $\sim$85 km s$^{-1}$, which is lower in brightness in comparison to the one of compact maser emission, it could be a result of the interior X-ray emitting plasma rather than by the cosmic rays accelerated by the SNR shocks. 

\item We compared Kes 69 with other interacting MM SNRs and concluded that Kes 69 offers an interesting case because of non-detection both RP and gamma-ray emission.

\end{enumerate}

\section*{Acknowledgments}

We would like to thank Dr. Aya Bamba for her valuable comments to improve this paper. We thank all the {\it Suzaku} team members for their support of the observation and software development. We also thank the referee for constructive comments and recommendations. AS is supported by the Scientific and Technological Research Council of Turkey (T\"{U}B\.{I}TAK) through the B\.{I}DEB-2219 fellowship program. TE thanks to the support of the Science Academy Young Scientists Program (BAGEP-2015). This work is supported in part by grant-in-aid from the Ministry of Education, Culture, Sports, Science, and Technology (MEXT) of Japan, No.15K05088(RY), No.18H01232(RY) and No.16K17702(YO).

$~$

Note added in proof: While this paper was under review, Nobukawa et al. (2018; arXiv:1801.07881), reported a discovery of the neutral iron line emission from Kes 69 and its flux is higher than the average of the GRXE, which agrees with our result.

$~$

{\it Facility}: {\it Suzaku} and  {\it Fermi}. 


\onecolumn

\twocolumn

\end{document}